\documentclass[12pt, letterpaper, oneside]{article}   
\usepackage{amsmath,amsfonts,amssymb,amsthm,bm,upgreek}
\usepackage{geometry}
\usepackage{cite}
\usepackage{mathtools}
\usepackage{stmaryrd}
\usepackage{bm}
\usepackage{multirow}
\usepackage{csvsimple}
\usepackage{setspace,dcolumn}
\usepackage{graphicx,psfrag,epsfig}
\usepackage{booktabs}
\usepackage{mhchem}
\usepackage{siunitx}
\usepackage{comment}
\usepackage{subcaption}
\usepackage[mathscr]{eucal}

\newcommand{\rmd}{\mathrm{d}}

\newcommand{\EEunits}[2]{\times10^{#1}\,\mathrm{#2}}
\newcommand{\units}[1]{\,\mathrm{#1}}
\newcommand{\eexp}[1]{\mathrm{e}^{#1}}

\newcommand{\kBT}{k_\mathrm{B}T}
\newcommand{\subm}[1]{_{\mathrm{#1}}}

\title{Physics-based Model for Nonuniform Thermionic Electron Emission from Polycrystalline Cathodes}
\author{
	Dongzheng Chen, Ryan Jacobs, and Dane Morgan\\
	Department of Materials Science and Engineering,
\\
	University of Wisconsin-Madison, \\
	1509 University Ave, Madison, \\
	WI 53706, USA
\\
	\and
	John Petillo \\
	Leidos, \\
	700 Technology Park Dr.,\\
	Billerica, MA 01821 USA \\
	\and
	Vasilios Vlahos \\
	Viasat Space Systems, \\
	2040 E. Technology Circle, \\
	Tempe, AZ 85284 \\
	\and
	Kevin L. Jensen \\
	Code 6362, MSTD, Naval Research Laboratory, \\
	Washington, DC 20375, USA \\
	\and
	John Booske \\
	Department of Electrical and Computer Engineering, \\
	University of Wisconsin-Madison, \\
	1415 Engineering Dr, \\
	Madison, WI 53706, USA \\
}
\begin{document}
\maketitle

\begin{abstract}
A new physics-based model is developed that predicts the emitted current from thermionic emission that accurately spans from the temperature-limited (TL) to the full-space-charge-limited (FSCL) regions. Experimental observations of thermionic electron emission demonstrate a smooth transition between TL and FSCL regions of the emitted-current-density-versus-temperature ($J-T$) (Miram) curve and the emitted-current-density-versus-voltage ($J-V$) curve. Knowledge of the temperature and shape of the TL-FSCL transition is important in evaluating the thermionic electron emission performance of cathodes, including predicting the lifetime. However, there have been no first-principles physics-based models that can predict the smooth TL-FSCL transition region for real thermionic cathodes without applying physically difficult to justify \textit{a priori} assumptions or empirical phenomenological equations. Previous work detailing the nonuniform thermionic emission found that the effects of 3-D space charge, patch fields (electrostatic potential nonuniformity on the cathode surface based on local work function values), and Schottky barrier lowering can lead to a smooth TL-FSCL transition region from a model thermionic cathode surface with a checkerboard spatial distribution of work function values. In this work, we construct a physics-based nonuniform emission model for commercial dispenser cathodes for the first time. This emission model is obtained by incorporating the cathode surface grain orientation via electron backscatter diffraction (EBSD) and the facet-orientation-specific work function values from density functional theory (DFT) calculations. The model enables construction of two-dimensional emitted current density maps of the cathode surface and corresponding $J-T$ and $J-V$ curves. The predicted emission curves show excellent agreement with experiment, not only in TL and FSCL regions but, crucially, also in the TL-FSCL transition region. This model provides a method to predict the thermionic emission from the microstructure of a commercial cathode, and improves the understanding on the relationship between thermionic emission and cathode microstructure, which is beneficial to the design of vacuum electronic devices.
\end{abstract}

\section{Introduction}
Thermionic electron emission cathodes based on porous, polycrystalline W combined with mixtures of metal oxides (typically $\mathrm{BaO-CaO-Al_2O_3}$) marked a significant evolutionary step in the history of thermionic cathodes, as these dispenser cathodes produce high-current-density emission with long lifetime due to their dynamically stable, low-work-function surfaces.\cite{Kirkwood2018} There are some widely-used mixture ratios for the metal oxides in the dispenser cathodes. The most common mix is $\mathrm{BaO:CaO:Al_2O_3=5:3:2}$ (a B-type cathode) produces emitted current densities of several $\mathrm{A/cm^2}$. There are other variations including the $4:1:1$ cathode (S-type), which is resistant to surface poisoning and can usually be operated at a temperature $30\units{^\circ C}$ lower than other types.\cite{Cronin1981,Vlahos2009,Gilmour2011,Jacobs2017} More recent dispenser cathodes include the M-type and scandate cathodes, which have a lower effective work function than the B- and S-type cathodes.\cite{Zhou2018,Liu2019,Wang2019} The B-, S- and M-type cathodes have constituted the majority of commercial thermionic cathodes for the past 50 years; used as the electron sources in numerous vacuum electronic devices (VEDs) such as communication devices, ion thrusters, thermionic energy converters, and free electron lasers. These applications, taken together, influence multiple facets of our modern life, ranging from defense, satellite communications, radar, and scientific research, to industrial-scale food production and manufacture of heat-harvesting renewable energy technology.\cite{Barker2005,Booske2008} 

Numerous experimental and computational studies have shown that the microstructure of real W-based cathodes is complex. The tungsten bodies are polycrystalline and porous, and the cathode surfaces are spatially heterogeneous, with the presence of machining marks from the cathode manufacturing process also contributing to the heterogeneity and causing local field enhancement effects.\cite{Gilmour2011,Jones1979,Jensen2003,Wan2012} One of the results of the complex microstructure is that W-based cathodes are spatially heterogeneous with a distribution of grain sizes and many types of exposed surfaces. These surfaces might have varied crystal facets and metal oxide coatings, each with an associated work function value, leading to highly non-uniform emission.\cite{Vlahos2009,Jacobs2017,Zhou2018,Forman1976Surface-studies,Haas1983,Norman1987Surface-structu,Vlahos2010} The non-uniform nature of thermionic electron emission from polycrystalline W has been observed experimentally by using thermionic electron emission microscopy (ThEEM).\cite{Norman1987Surface-structu,Haas1967,Mroz2019a,Tuck1979,Wan2012Scandium-oxide-,Vaughn2009,Vaughn2010,Kordesch2013,Wan2013,Ren2017,Mroz2018} In a representative ThEEM image, at a particular temperature, certain grains of the W surface are bright while others remain dark, indicating that some grains are more emissive than others, due to factors such as lower work function, surface topography, etc.

Emitted-current-density-versus-temperature, or $J-T$ (Miram) curves and emitted-current-density-versus-voltage, or $J-V$ ($I–V$) curves are commonly used to evaluate the cathode performance. Both the $J–T$ and $J–V$ curves of a cathode can be divided into three regions: temperature-limited (TL) region, full-space-charge-limited (FSCL) region, and the TL-FSCL transition region. The TL region is in the low-temperature end of an $J–T$ curve or the high-voltage end of an $J–V$ curve. Its behavior can be well described with the Richardson–Laue–Dushman equation\cite{Richardson1922,Dushman1930a} with Schottky barrier lowering\cite{Schottky1923}. The FSCL region is in the high-temperature end of a $J–T$ curve or the low-voltage end of a $J–V$ curve. The behavior can be predicted by the Child–Langmuir law\cite{Child1911,Langmuir1923The-Effect-of-S} and Langmuir and Fry’s studies\cite{Langmuir1923The-Effect-of-S,Fry1921}, including provision for two-dimensional edge-correction effects\cite{Luginsland1996,Lau2001,Umstattd2001,Luginsland2002Beyond-the-Chil,Quan2009a,Sitek2021,Sitek2021a}. Experimental observations on real thermionic cathodes show that the TL-FSCL region is usually smooth, sometimes referred to as the “roll-off”. Despite this seemingly simple observed behavior, it has remained an ongoing challenge to develop a physics-based emission model which is able to accurately predict the behavior of both $J–T$ and $J–V$ curves from polycrystalline cathodes over the entire operational domain of temperature and anode-cathode voltage, and especially challenging to capture the smooth transition between the TL and FSCL regions for real cathodes. Thermionic cathodes are typically operated on the FSCL side near the TL-FSCL transition region, so that the changes in cathode temperature over time do not cause large variations to the emitted current and that the emission is stable over the predicted lifetime of the device.

Some empirical descriptions on the smooth TL-FSCL region have been developed, including the empirical Longo–Vaughan equation\cite{Longo1980,Vaughan1986A-synthesis-of-}, a continuous Gaussian distribution of work function\cite{Gilmour1994}, the work function distribution mathematical treatment of emission data\cite{Tonnerre1983}, and the practical work function distribution function (PWFD)\cite{Cattelino1997}. However, all of these models are based on empirical equations or difficult-to-justify a priori assumptions, for example, the assumption that different work function patches do not interact. Furthermore, these empirical descriptions are not able to reveal the fundamental origin of the smooth behavior of the TL-FSCL transition, thus limiting their usefulness for modeling cathode behavior under different operating conditions.

A number of previous works have studied the interplay of a heterogeneous cathode surface on the resulting thermionic emission, and have sought to connect the smooth TL-FSCL transition to the spatial distribution of work function values. The theory of the anomalous Schottky effect\cite{Hansen1966} studied the contribution of the patch field effect (electrostatic potential nonuniformity on the cathode surface based on local work function values) and the Schottky barrier lowering effect on the smoothness of the TL-FSCL transition in $J-V$ curves. Studies on space charge effects\cite{Sitek2021,Sitek2021a,Chernin2020,Jassem2021} reveal the contribution of 3-D space charge fields on the smooth transition in $J-T$ curves. However, the TL-FSCL transition behaviors predicted from these two separate sets of studies are sharper than experimental observations, indicating that some physical effects are missing. There has been no physics-based emission model which can predict the TL-FSCL transition in agreement with experimental results, although Longo and Vaughan speculated\cite{Longo1980,Vaughan1986A-synthesis-of-} that sharper Miram curve knees might be associated with more uniform work function surfaces, or “better” cathodes. Our recent work\cite{Chen2021} developed a physics-based model that included the effects of nonuniform thermionic emission, 3-D space charge, patch fields, and Schottky barrier lowering. This work gives a mathematical method to calculate the emitted current from a cathode with a spatially heterogeneous work function distribution in a parallel diode, and is able to predict a smooth and gradual TL-FSCL transition comparable with experimental observations by using a checkerboard work function distribution. These findings were encouraging, and indicated our model may be successful in predicting the emission of a real cathode, including the smooth TL-FSCL transition, by applying a two-dimensional work function map obtained from the same real cathode. 

In this work, we construct a two-dimensional work function map by incorporating the grain orientation via electron backscatter diffraction (EBSD) and the facet-orientation-specific work function values from density functional theory (DFT) calculations. We use this work function map in conjunction with the nonuniform emission model developed in our previous work\cite{Chen2021} to predict both the $J–T$ (Miram) and $J–V$ ($I–V$) curves, including the TL-FSCL transition. Overall, we find semi-quantitative agreement of our predicted results with experimental measurements. This is the first time a physics-based thermionic emission model incorporating heterogeneous surface effects from a work function distribution on a real commercial thermionic cathode has been used to successfully model the experimental emission over a wide domain of temperature and applied voltage. 

\section{Methods}
\subsection{Cathode sample\label{sec:cathode_sample}}
The cathode analyzed in this work is a commercial S-type cathode made by 3M Technical Ceramics. The cathode was made of $80\%$ density W using standard manufacturing methods and impregnated with an oxide mixture of $\mathrm{BaO:CaO:Al_2O_3=4:1:1}$. The cathode was cylinder-shaped with a $2.77\units{mm}$ diameter and $0.97\units{mm}$ height, as measured after the emission test.

\subsection{Emission measurement}
\begin{figure}[htbp]
	\centering
	\includegraphics[scale=1]{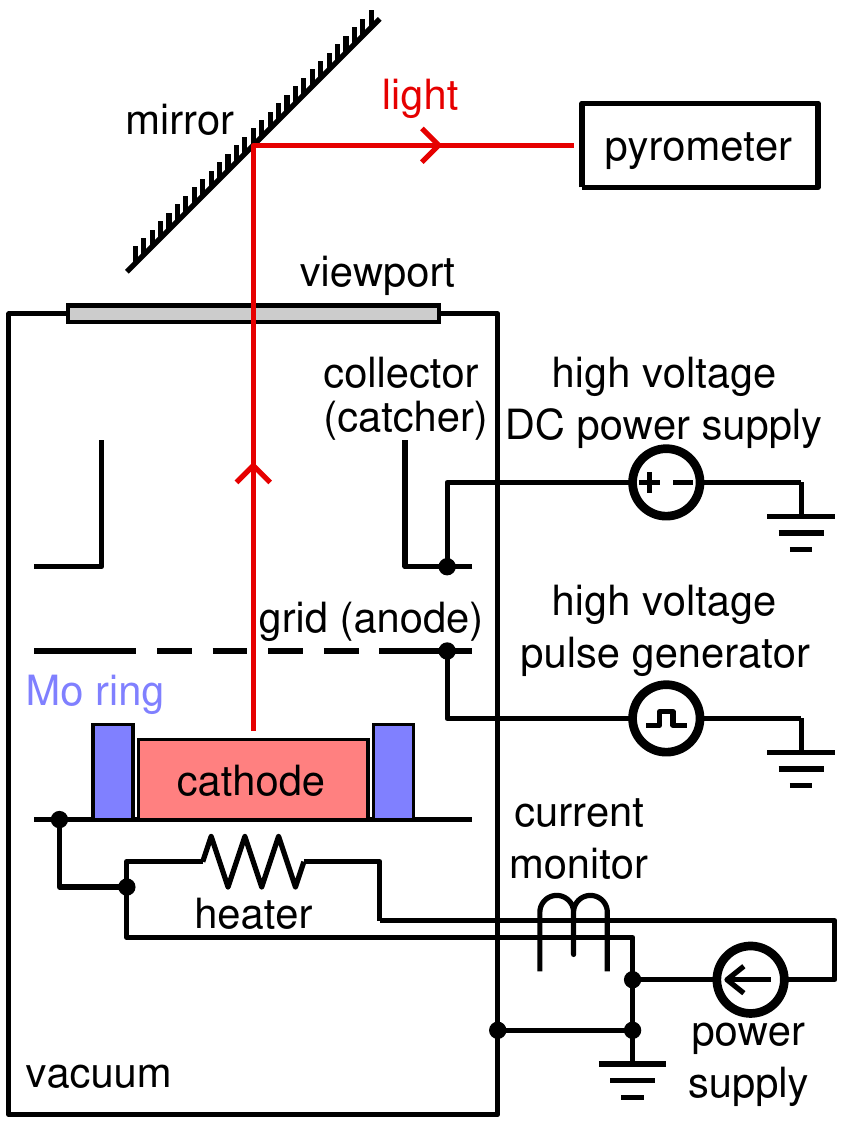}
	\caption{\label{p2f1}
		Sketch of the closely spaced diode testing vehicle used to measure the thermionic emitted current. The rectangle filled with pink is the cathode. The purple rectangles around the cathode represent the molybdenum ring used to shield the side emission.
	}
\end{figure}

The experimental results of emitted current were measured in a closely spaced diode testing vehicle (Fig. \ref{p2f1}). The heater and the anode fixtures were manufactured by L3-Harris. The anode-cathode distance in this setup for a $0.97\units{mm}$ high cathode was designed to be $d=1.06\units{mm}$. A molybdenum ring was placed around the cathode to shield the emission from the sides. The height of the molybdenum ring was $1.14\units{mm}$, which was $0.17\units{mm}$ higher than the cathode. The inner diameter (ID) $2.90\units{mm}$ was $0.13\units{mm}$ larger than the diameter of the cathode. The heater filament was powered by a Keithley 2200-20-5 programmable power supply, and was operated under constant current mode.

To make it possible to measure the temperature of the cathode surfaces using a pyrometer during operation, a triode design was used with a cylinder as the current collector, or “catcher”. The temperature of the cathode surface was measured with a Cat. No. 8622 optical pyrometer made by Leeds \& Northrup Co., which is a $\lambda=0.65\units{\mu m}$ single-wavelength disappearing filament pyrometer. The electron emission cathode industry often simply uses the pyrometer reading to indicate the cathode temperature, reporting it as the brightness temperature. However, the true temperature of the cathode surface is needed to use our nonuniform emission model\cite{Chen2021}. We calibrated the temperature values using Planck’s law. The radiation of the cathode received by the disappearing pyrometer at wavelength $\lambda$ is:
\begin{equation}\label{p2e1}
	tr\epsilon\frac{2hc^2}{\lambda^5}\frac{1}{\eexp{hc/(\lambda\kBT)}-1}=\frac{2hc^2}{\lambda^5}\frac{1}{\eexp{hc/(\lambda\kBT\subm{b})}-1}
\end{equation}
where $T\subm{b}$ is the pyrometer reading (brightness temperature), $T$ is the calibrated “true” temperature of the cathode surface to be used in the emission model, $k$ is the Boltzmann constant, $h$ is the Planck constant, $c$ is the speed of light. The values of the transmissivity of the viewport $t=0.93$ and the reflectivity of the mirror $r=0.76$ in the optical path were as measured. In this study, we used the emissivity value of $\epsilon=0.52$ recommended for impregnated W cathodes.\cite{Cronin1981} The uncertainty of the measured temperature values was approximately $\pm20\units{^\circ C}$.

The cathode was activated before emission test, following the instructions recommended by the cathode manufacturer, 3M Technical Ceramics. The activation process includes four steps: (1) Slowly increase the cathode temperature to a brightness temperature of $1000\units{^\circ C}$, and then hold this temperature for 30 minutes. (2) Continue to increase the cathode temperature to a brightness temperature $1175--1200\units{^\circ C}$ and hold for 1 hour. (3) Cool the cathode to a brightness temperature $1100--1150\units{^\circ C}$ and hold for 2 hours. (4) Reduce cathode temperature and measure the emitted current while cooling down the cathode. The pressure was kept below $5\EEunits{-6}{torr}$ during the activation process. 

During the emission measurements, the grid was biased with a PVX-4110 high voltage pulse generator made by Directed Energy (DEI), which was powered by a DC high voltage power supply made by Glassman High Voltage, Inc. with Model No. PS/ER02R150-115, and controlled by a low voltage pulse generator Model 575 pulse/delay generator made by Berkeley Nucleonics Corp (BNC). The catcher was biased with a DC high voltage power supply made by Glassman High Voltage, Inc. with Model No. PS/EQ005R240-22 and was kept more positively biased than the grid. The voltages of the grid and the catcher were measured with a LeCroy 44Xs oscilloscope. The emitted current was measured with the same oscilloscope via a Model 4100C current monitor made by Pearson Electronics, Inc.

\subsection{Microstructure characterization}
The cathode surface grain orientation was characterized using electron backscatter diffraction (EBSD) in a FEI Helios G4 Plasma FIB/FESEM/EBSD/EDS workstation after the emission test. The surface of commercial dispenser cathodes are usually rough due to the machining process of cutting the cathode pellets on a lathe. The machining typically produces micrometer-scale ridges, and those differences can be seen in variation in emission properties.\cite{Jensen2006a} Confidence index (CI) values in the EBSD results were used to quantify the likelihood of correct grain orientation labeling.\cite{Field1997} CI standardization, one of the built-in clean-up algorithms in OIM Analysis™ by EDAX, a software for EBSD analysis, was used to process the raw EBSD data. Pixels with low CI values after applying the CI standardization clean-up procedure were considered as areas where grain orientations were unable to be correctly labeled by EBSD\cite{Nowell2005} and the surface facet orientation could not be reliably determined. Visual inspection showed that the majority of pixels with CI lower than 0.1 were associated with nonemitting areas, such as rough valleys, depressions, grain boundaries, and pores.\cite{Wright2006} Thus, areas with $\mathrm{CI}<0.1$ were considered as no-emit areas, and the grain orientation of areas with $\mathrm{CI}\geq0.1$ were considered as recognizable areas. We then used simulations described below to establish a two-dimensional work function map $\phi\subm{DFT}(x,y)$ for $\mathrm{CI}\geq0.1$ areas.

\subsection{Density functional theory work function values}
\begin{figure}[htbp]
	\centering
	\includegraphics[scale=1]{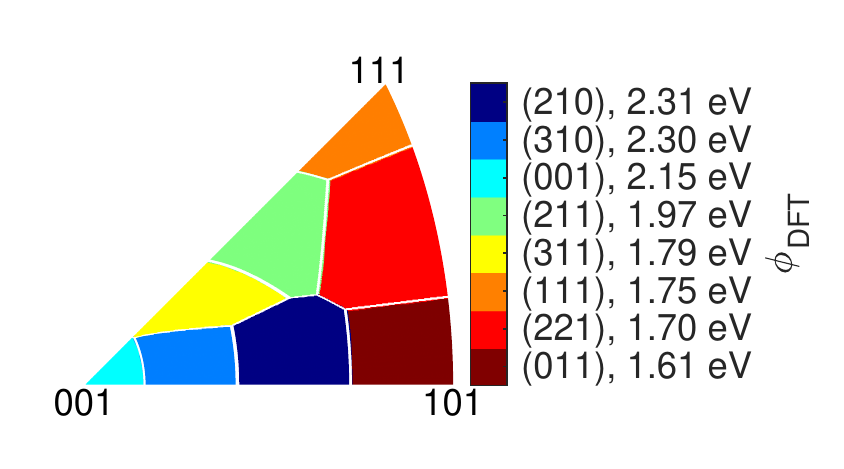}
	\caption{\label{p2f2}
		Inverse polar figure (IPF) showing how a grain orientation is grouped into one of the eight orientation groups using the nearest neighbor algorithm. The colors indicate the work function values assigned to each orientation group, which are the work function values of the W surface with most stable stoichiometric Ba-O adsorption calculated by density functional theory (DFT).
	}
\end{figure}

Previous density functional theory (DFT) studies have calculated the work functions and surface stabilities of tungsten surfaces with Ba, O, and Ba-O adsorbates of eight different orientations: $(001)$, $(011)$, $(111)$, $(210)$, $(211)$, $(221)$, $(310)$, $(311)$.\cite{Jacobs2017,Zhou2018,Vlahos2010} Auger analysis indicates that the active state for impregnated cathodes can be reproduced by a near monolayer of the stoichiometric Ba-O on the W surface.\cite{Haas1983} Only the DFT work function value for the most stable stoichiometric Ba-O adsorption are used to assigned to each orientation (Table \ref{p2t1}). For a high-index orientation $(hkl)$ other than the calculated eight orientations, the nearest neighbor algorithm is used to predict its work function.\cite{Chen2019} It is assumed that $(hkl)$ orientation has the same work function as the one among the calculated eight orientations with the smallest misorientation with $(hkl)$ (Fig. \ref{p2f2}).

A few studies have estimated the uncertainty of DFT work function values by comparing DFT results with experimental results. De Waele \textit{et al.}\cite{DeWaele2016} compared the experimental work function values for different surface orientations for a number of metals with the values predicted by the Perdew-Burke-Ernzerhof parametrization of the generalized gradient approximation (PBE-GGA) method. They did a linear fit of DFT values $\phi\subm{DFT}$ and experimental values $\phi\subm{exp}$. The result was the equation $\phi\subm{exp}=\beta_1\phi\subm{DFT}+\beta_0$, where the values of the fitted coefficients were $\beta_1=0.99\pm0.02$ and $\beta_0=0.30\pm0.09\units{(eV)}$. Tran \textit{et al.}\cite{Tran2019} also compared their DFT results, $\phi\subm{DFT}$, with experimental values, $\phi\subm{exp}$, on single crystals. They made a single-parameter $\phi\subm{DFT}=\phi\subm{exp}-c$ least square fit, where their result was $c=0.30\units{eV}$. Both results indicate that DFT work function predictions of metals using GGA-level functionals tend to underestimate the work function values by approximately $0.30\units{eV}$, on average, compared with experimental results, and that the error of the estimate is on the scale of tenths of eV even after the linear fit. Due to this known underestimation, we consider the shift between experimental and calculated work function as a fitting parameter in our emission modeling (more details in Section \ref{sec:emission_modeling}).  

\subsection{Emission modeling\label{sec:emission_modeling}}
It is prohibitively difficult to accurately measure, and thus, to know, the actual anode-cathode distance $d$ at the operating temperatures in our test fixture. Therefore, to better compare the results of the emission model with the experimental results, in this work, we obtained the effective anode-cathode distance $d$ by fitting the FSCL data points with the Child-Langmuir law with the finite temperature correction\cite{Langmuir1923The-Effect-of-S}:
\begin{equation}\label{p2e2}
	J\subm{FSCL}=\frac{4\epsilon_0}{9}\sqrt{\frac{2e}{m}}\frac{(V-V\subm{m})^{3/2}}{(d-z\subm{m})^2}
	\frac{9}{8\sqrt{\pi}}\eta^{-3/2}\left(\int_0^\eta\frac{\rmd\eta}{\sqrt{\mathrm{erfx\,}\sqrt{\eta}-1+2\sqrt{\eta/\pi}}}\right)
\end{equation}
where $\epsilon_0$ is the vacuum permittivity, $e$ is the elementary charge, $m$ is the electron mass, $V$ is the anode-cathode voltage, $d$ is the anode-cathode distance, $V\subm{m}$ and $z\subm{m}$ are the voltage and the location from cathode of the voltage minimum, $\eta=e(V-V\subm{m})/(\kBT)$ where $k$ is the Boltzmann constant and $T$ is the temperature, and $\mathrm{erfcx}$ is the scaled complementary error function. Instead of using the as-designed value of the anode-cathode distance, we used the fitted value for the emission model, which we believe is a more accurate value for high temperatures during emission measurements.

In the theory of the Child-Langmuir law with the finite temperature correction on a uniform cathode,\cite{Langmuir1923The-Effect-of-S} the voltage minimum satisfies the Richardson-Laue-Dushman equation: $J=AT^2\eexp{-eV\subm{m}/(\kBT)}$, while the position of the voltage minimum $z\subm{m}=0$ at the TL-FSCL transition. When fitting $d$ using Eq. \ref{p2e2}, we made the same assumptions: $z\subm{m}=0$ and $J=AT^2\eexp{-eV\subm{m}/(\kBT)}$, where $J$ is the emitted current density for which we used experimental results. Here, $A=4\pi mek^2/h^3=120.173\units{A\,cm^{-2}\,K^{-2}}$ is the Richardson constant where h is Planck’s constant.

It is not practical to do EBSD on the whole surface of a cathode, so we characterized the grain orientation on a representative area of the cathode surface (more details in Section \ref{sec:sensitivity_analysis}), and used periodic boundary conditions on the edges of the work function map, considering that the nonuniform emission model\cite{Chen2021} was designed for spatially periodic work function maps. Considering the error in the DFT work function value, we added a constant shift $\Delta\phi$ on the DFT work function values, $\phi\subm{DFT}(x,y)$, to get a shifted work function map, $\phi(x,y)=\phi\subm{DFT}(x,y)+\Delta\phi$, for the $\mathrm{CI}\geq0.1$ areas.

The roughness of the thermionic cathode used in this study is mainly due to the machining and the grain structures. The range of the field enhancement factor values expected from the roughness features of typical thermionic cathodes is usually small, with an estimated upper bounds of $5$.\cite{Jensen2003,Jensen2006a,Miller2007,Miller2009} Even in the case that the applied electric field is $500\units{V/mm}$, the difference in the Schottky barrier lowering between a surface with a field enhancement factor of $\beta=5$ and a perfectly flat surface $\beta=1$ is only $0.033\units{eV}$, which will add a negligible enhancement to the thermionic emission, compared with the uncertainty of DFT work function values. Therefore, for simplicity in our model, we assumed the cathode surface was perfectly flat and therefore neglected field enhancement effects.

The grain orientation of areas with $\mathrm{CI}\geq0.1$ were considered as recognizable areas, and a work function map $\phi(x,y)=\phi\subm{DFT}(x,y)+\Delta\phi$ was assigned to these areas. If we let the cathode Fermi level be zero, then the boundary condition of the Poisson’s equation for $\mathrm{CI}\geq0.1$ areas is its vacuum level $V(x,y,z=0)=-\phi(x,y)/e$.\cite{Chen2021}

As the majority of $\mathrm{CI}<0.1$ pixels were associated with nonemitting areas, such as rough valleys, depressions, grain boundaries, and pores,\cite{Wright2006} we obtain the boundary condition for the cathode surface $V(x,y,z=0)$ for the $\mathrm{CI}<0.1$ areas by solving the 2-D Laplace’s equation $\nabla^2V(x,y)=0$ where the boundary conditions are the values of $\mathrm{CI}\geq0.1$ areas. In this way, we obtain the boundary condition of the whole cathode surface, for both $\mathrm{CI}\geq0.1$ areas and $\mathrm{CI}<0.1$ areas, which will be used as the input of the nonuniform emission model\cite{Chen2021}. There is only one fitting parameter $\Delta\phi$ in this model. 

In the model, the potential energy for an electron present in the space within the diode is obtained by solving Poisson’s equation, where the charge density is a nonlinear function of the potential energy in the space. The effect of Schottky barrier lowering is included when calculating the potential energy. The patch field effect is naturally included in the non-equipotential boundary condition at cathode surface $V(x,y,z=0)$, and 3-D Poisson’s equation includes the 3-D space charge effect. Therefore, such a nonuniform emission model includes the effects of 3-D space charge, patch fields, and Schottky barrier lowering, but neglects the effects of the lateral motion of electrons and the quantum effects (e.g., electron tunneling). More information on the physics and specific calculation methodology of our nonuniform emission model can be found in Ref. \cite{Chen2021}.

\section{Results and Discussion}
\subsection{Spatial distribution of work function}
\begin{figure}[htbp]
	\centering
	\includegraphics[scale=1]{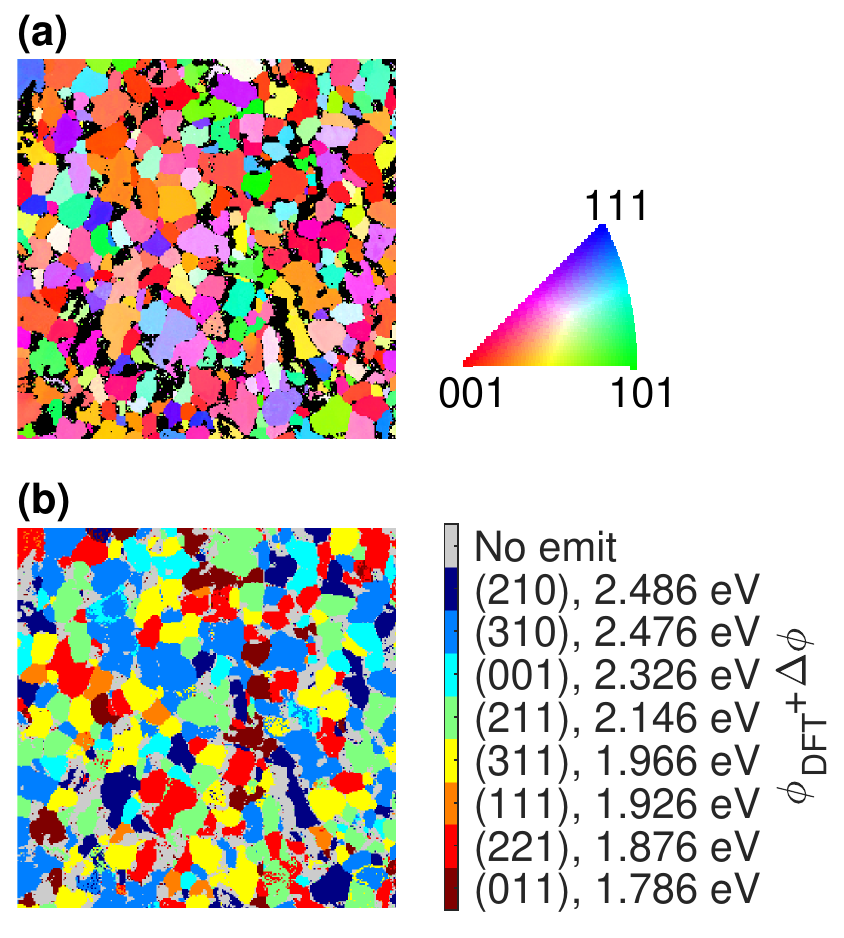}
	\caption{\label{p2f3}
		(a) Electron backscatter diffraction (EBSD) inverse polar figure (IPF) of a commercial S-type cathode after clean-up. The areas where it is considered that grain orientations are unrecognized by EBSD are plotted in black. (b) Work function map by assigning the density functional theory (DFT) work function value [5] with a shift of $\Delta\phi=0.176\units{eV}$ to the grain orientation map (a) after grouping the orientations into one of the eight orientation groups.
	}
\end{figure}

The spatial distribution of grain orientation was characterized using EBSD after the emission testing was concluded. Emission testing and grain orientation analysis were performed for the same cathode samples, and were performed after emission testing to ensure that any microstructural evolution that may have occurred during the high temperature activation and emission testing processes was captured. Fig. \ref{p2f3}a shows the two-dimensional maps of grain orientation of a representative portion of a cathode surface (more details in Section \ref{sec:sensitivity_analysis}). The percentage of each orientation group in the map is listed in Table \ref{p2t1}. 

We measured the emitted current from a commercial S-type cathode made by 3M Technical Ceramics (Section \ref{sec:cathode_sample}) for various anode-cathode voltages and temperatures (Fig. \ref{p2f4}). The anode-cathode distance was obtained by fitting the $24$ data points above $1340\units{K}$ in Fig. \ref{p2f4}a using Eq. \ref{p2e2}, and the result was $d=1.132\units{mm}$. This is close to the designed value of $1.06\units{mm}$. We ascribe the discrepancy between the fitted distance and the designed value to several reasonable factors that include a likely small difference between the designed distance and the actual fabricated distance (at room temperature) as well as the effects of electron optics and thermal expansion. The constant work function shift was obtained by fitting all of the data points in Fig. \ref{p2f4}a with the nonuniform emission model, and the result was $\Delta\phi=0.176\units{eV}$, which indicates that DFT underestimated the work function values compared with the thermionic emission test results. This result is consistent with previous studies on the error of DFT work function values\cite{DeWaele2016,Tran2019} in both the sign and magnitude of the error (underestimation by DFT of about $0.3\units{eV}$). Fig. \ref{p2f3}b is the predicted work function map, obtained by applying shifted DFT work function values to the grain orientation map (Fig. \ref{p2f3}a).

\begin{table}[hbtp]
	\caption{\label{p2t1} List of the eight orientations with work function values predicted using density functional theory (DFT). $\phi\subm{DFT}$ is the DFT work function value for the most stable stoichiometric Ba-O adsorption for each orientation. $\phi\subm{DFT}+\Delta\phi$ is the shifted work function value where the shift is $\Delta\phi=0.176\units{eV}$. The “Percentage” column shows the percentage of each orientation group in Fig. \ref{p2f3}.}
	\centering
	\begin{tabular}{cccc}
		\toprule
		Orientation & $\phi\subm{DFT}\units{(eV)}$ & $\phi\subm{DFT}+\Delta\phi\units{(eV)}$ & Percentage \\
		\midrule
		$(001)$ & $2.15$ & $2.326$ & $6.3\%$
\\
		$(011)$ & $1.61$ & $1.786$ & $5.5\%$ \\
		$(111)$ & $1.75$ & $1.926$ & $2.3\%$ \\
		$(210)$ & $2.31$ & $2.486$ & $8.9\%$ \\
		$(211)$ & $1.97$ & $2.146$ & $14.0\%$ \\
		$(221)$ & $1.70$ & $1.876$ & $12.7\%$ \\
		$(310)$ & $2.30$ & $2.476$ & $19.3\%$ \\
		$(311)$ & $1.79$ & $1.966$ & $13.8\%$ \\
		Unrecognized & - & - & $17.1\%$ \\
		\bottomrule
	\end{tabular}
\end{table}

\subsection{Emitted current density}
\begin{figure}[htbp]
	\centering
	\includegraphics[scale=0.95]{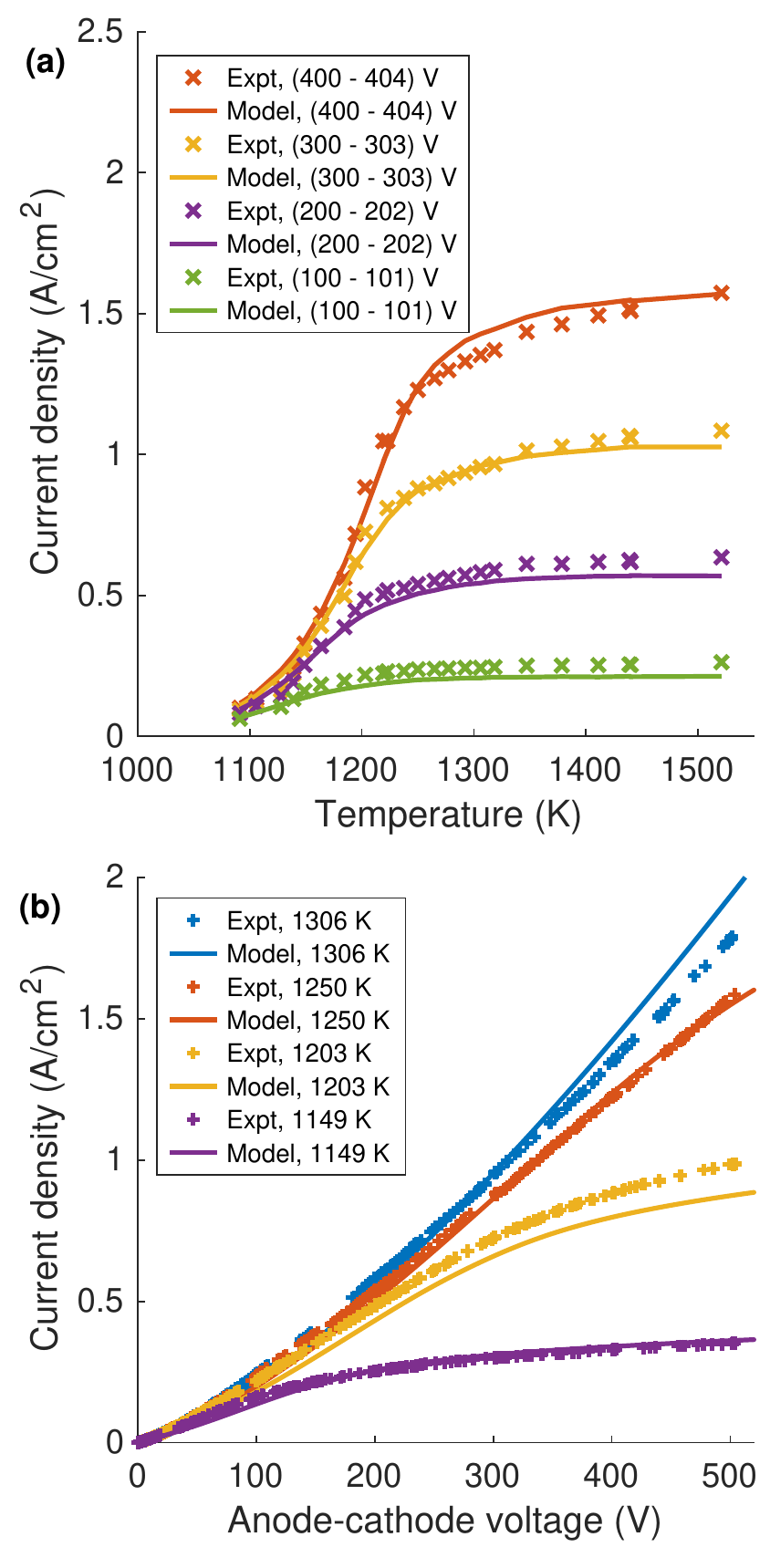}
	\caption{\label{p2f4}
		Experimental data ($\times$ and $+$ symbols) of an S-type cathode compared with the emitted current density predicted with nonuniform emission model (lines) at different anode-cathode voltages. (a) $J-T$ curves for different anode-cathode voltages $V$. The measured $V$ values for the data points of red $\times$ symbols are between $400\units{V}$ and $404\units{V}$, and $300\units{V}\leq V\leq303\units{V}$ for yellow, $200\units{V}\leq V\leq202\units{V}$ for purple, $100\units{V}\leq V\leq101\units{V}$ for green. (b) $J-V$ curves at different temperatures.
	}
\end{figure}

Fig. \ref{p2f4} shows the experimental emission data from the S-type cathode and the emitted current density predicted by applying the nonuniform emission model \cite{Chen2021} to the work function map (Fig. \ref{p2f3}b). The predicted TL-FSCL transition regions are as smooth as the experimental observations for both the $J-T$ and $J-V$ curves, resulting in semi-quantitative agreement between our model and experimental measurements.

There are only two fitting parameters in our model: the anode-cathode distance $d=1.132\units{mm}$ and the constant shift on DFT work function values $\Delta\phi=0.176\units{eV}$. The main effect of a different $d$ is to scale up and down the FSCL current, while the effect of a different $\Delta\phi$ is to scale up and down the TL current or shift the TL region to a lower or higher temperature in $J-T$ curves. Using fitted values for these two parameters helps to get a better fit for the TL and FSCL regions, enabling a better comparison on the TL-FSCL transition regions between predicted curves and experimental results. The exact values of both of these fitted parameters have negligible effects on the shape of the TL-FSCL transition region.

The smooth TL-FSCL transition in the predicted curves arises as a natural consequence of the nonuniform emission from the polycrystalline cathode with a nonuniform spatial distribution of work function. Previous studies\cite{Sitek2021,Sitek2021a,Chernin2020,Jassem2021} show that the 3-D space charge effect plays a significant role in making the transition region smooth for a nonuniform cathode. However, when using a work function map for a real cathode derived from DFT and EBSD, a model only considering the 3-D space charge effect predicts a TL-FSCL transition in a Miram curve sharper than experimental results.\cite{Chernin2020} The nonuniform emission model used in this work\cite{Chen2021} includes all of the effects of 3-D space charge, patch fields, and Schottky barrier lowering. This result shows that including all of these effects is required to predict the smooth TL-FSCL transition region in the $J-T$ and $J-V$ curves, not only for the checkerboard model cathode illustrated in\cite{Chen2021}, but also in a work function map of a real cathode.

The Richardson constant $A$ is an important factor in the Richardson-Laue-Dushman equation. Its theoretical value is $A=4\pi mek^2/h^3=120.173\units{A\,cm^{-2}\,K^{-2}}$. In multiple previous studies, the Richardson constant was experimentally obtained by fitting both the Richardson constant and the effective work function in the Richardson-Laue-Dushman equation to the experimental emission data under the assumption that the cathode is uniform and has a single work function value.\cite{Fomenko1966} It has been observed that, using this method, the experimental values of the Richardson constant differ from the theoretical value, sometimes by many orders of magnitude.\cite{Gilmour1994,Fomenko1966} However, the Richardson constant does not need to be fit in our model, and is assumed to be fixed to its theoretical value. The agreement between our experimental and predicted $J-T$ and $J-V$ curves indicates that the alteration of the Richardson constant is not needed here. Thus, a key strength of our present model is that knowledge of the fractions of different surface terminations present, their arrangement in 2D space on the surface, and their work functions are all that is required for the nonuniform emission model to provide a physically complete picture of the emission.

\subsection{Two-dimensional emission map}
\begin{figure}[htbp]
	\centering
	\includegraphics[scale=1]{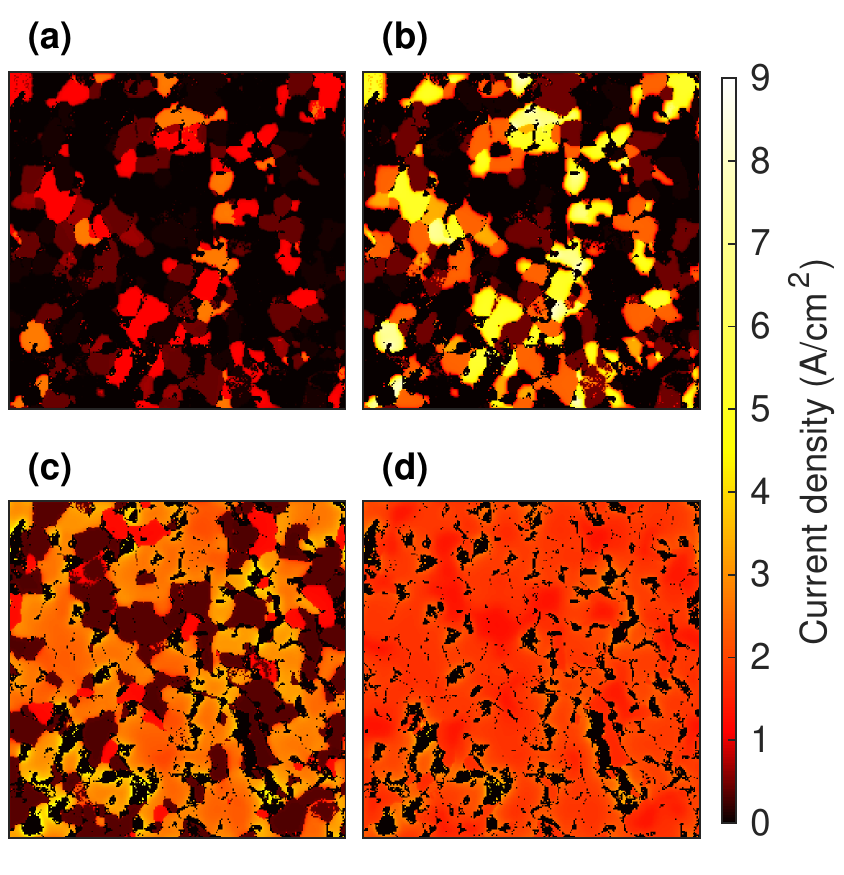}
	\caption{\label{p2f5}
		Emitted current density maps predicted using nonuniform emission model for a cathode with work function map as Fig. \ref{p2f3}b at anode-cathode voltage $V = 400\units{V}$ and distance $d=1.132\units{mm}$, at different temperatures: (a) TL region: temperature $T=1149\units{K}$, average emitted current density $J = 0.340 \units{A/cm^2}$, (b) transition region: $1250 \units{K}$, $J = 1.234 \units{A/cm^2}$, (c) transition region but with an average emitted current density close to the full-space-charge-limited (FSCL) value: $1411 \units{K}$, $J = 1.525 \units{A/cm^2}$, (d) FSCL region: $1521 \units{K}$, $J = 1.552 \units{A/cm^2}$.
	}
\end{figure}

\begin{figure}[htbp]
	\centering
	\includegraphics[scale=1]{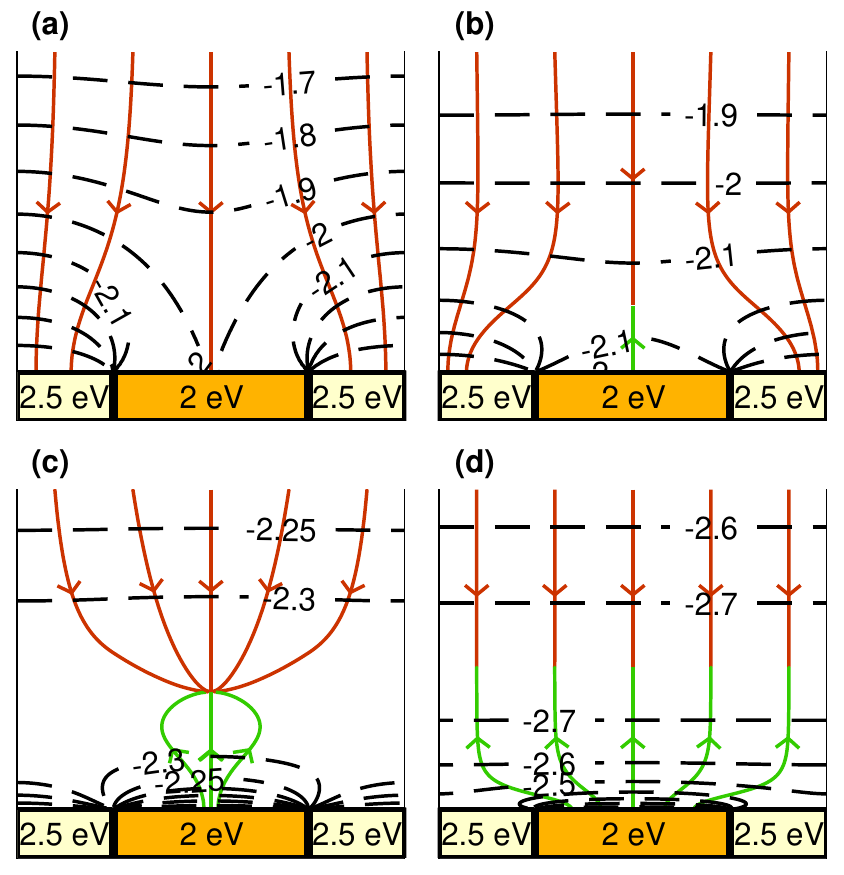}
	\caption{\label{p2f6}
		Schematic figures illustrating the effects of patch fields and 3-D space charge in different regions: (a) temperature-limited (TL) region, (b) transition region, (c) transition region with an average emitted current density close to full-space-charge-limited (FSCL) value, (d) full-space-charge-limited (FSCL) region. The anode (not shown in the figures) is far away on the top of each subfigure, and the cathode is on the bottom of each subfigure, with a low work function patch ($2 \units{eV}$) surrounded by high work function patches ($2.5 \units{eV}$). Dashed black curves are the equipotential curve of the electrostatic potential (unit: V). The red and green solid curves are the electric flux lines. The red ones are for the electric flux lines starting from the anode while the green ones for those starting from the cathode. The aspect ratio of different subfigures may be different, and may not be $1:1$, so the electric flux lines may appear not perpendicular to the equipotential curves.
	}
\end{figure}

Fig. \ref{p2f5} shows how the calculated emitted current density maps change as temperature increases and the emission changes from the temperature-limited (TL) region (Fig. \ref{p2f5}a), to the transition region (Fig. \ref{p2f5}b and \ref{p2f5}c), and finally to the full-space-charge-limited (FSCL) region (Fig. \ref{p2f5}d). To better illustrate the effects of the patch fields and space charge, we plotted schematic figures of equipotential curves and electric flux lines in the space in front of a low work function patch surrounded by high work function patches in TL, transition, and FSCL regions (Fig. \ref{p2f6}).

In the TL region (Fig. \ref{p2f5}a), the space charge effect is negligible, and therefore the low work function patches emit more than the high work function patches. This result that the emitted current density varies across different grains due to the difference in their work function values is consistent with experimental thermionic electron emission microscopy (ThEEM) images obtained in the TL region.\cite{Norman1987Surface-structu,Haas1967,Mroz2019a,Tuck1979,Wan2012Scandium-oxide-,Vaughn2009,Vaughn2010,Kordesch2013,Wan2013,Ren2017,Mroz2018}

As the schematic figures show, in the TL (Fig. \ref{p2f6}a) and the transition region (Fig. \ref{p2f6}b), the low work function patch faces a voltage minimum lower than its surface (the local vacuum level), especially at the patch edges, due to the patch field effects from its neighboring high work function patches. Therefore, the local emitted current density from the edge of low work function patches is smaller than the center of the patches (Fig. \ref{p2f5}a and \ref{p2f5}b). This result is different from the edge effect in a cathode surface with nonuniform emission but without patch field effects, where the edges of high-emitting patches emit more than the center of the patches due to the low space charge in front of their neighboring low-emitting patches.\cite{Umstattd2001,Luginsland2002Beyond-the-Chil,Sitek2021a}

In the transition and FSCL regions (Fig. \ref{p2f6}b, \ref{p2f6}c, and \ref{p2f6}d), the low work function patches tend to have more significant space charge effects than the high work function patches due to its higher local emitted current density, and will have a voltage minimum in front of its surface at a lower temperature. Such space charge effect causes that the emission from the low-emitting patches continues to increase while the high-emitting patches start to emit less, and that the emitted current density tends to be increasingly uniform due to the 3-D space charge effect, as the temperature increases from the transition region to the FSCL region (Fig. \ref{p2f5}b, \ref{p2f5}c, and \ref{p2f5}d).

In our nonuniform emission model, even though electrons are restricted along the cathode-anode direction with no lateral momentum, our model is able to predict the trend of the change in the emission nonuniformity as temperature changes. Such a trend has also been observed in experiments\cite{Li2006a,Li2007} and is consistent with some previous computational studies\cite{Sitek2021,Sitek2021a,Chernin2020,Jassem2021}.

\subsection{Sensitivity analysis and sources of error\label{sec:sensitivity_analysis}}
It is computationally expensive to simulate large areas (for example, $0.1 \units{mm^2}$ or larger) with the nonuniform emission model and time-consuming to characterize the grain orientation of a large area where there are a large number of grains. The computational cost is significantly increased in beam optics simulations where millions of time steps are typically used.\cite{Petillo2002The-michelle-th} To determine the representativeness of the statistics of the surface facets and ascertain the relationship between the uncertainty of the predicted emitted current and the size of the work function map, we characterized a total of 9 EBSD maps on different regions of the S-type cathode, for a total examined area of $0.15 \units{mm^2}$, and calculated the resulting emitted current density as a function of the examined area of the cathode surface.

To evaluate the effect of the uncertainty in the work function values to the predicted emitted current density, we calculated the emitted current density from a work function map by applying $\phi(hkl)=\phi\subm{DFT} (hkl)+ \Delta\phi(hkl)$ to the grain orientation map in Fig. \ref{p2f3}a, where the $(hkl)$ is one of the eight grain orientations assigned with DFT work function values, and the work function shift for the eight grain orientations $\Delta\phi(hkl)$ are assumed to be independent and identically distributed (i.i.d.) following the normal distribution $\Delta\phi(hkl)\sim N(0.176\units{eV},\sigma\subm{DFT}^2)$, where the standard deviation $\sigma\subm{DFT}$ represents a phenomenological error in our DFT measurements. We have generated 2500 random work function maps for each $\sigma\subm{DFT}$ value and calculated the variability of their emitted current densities.

\begin{figure}[htbp]
	\centering
	\includegraphics[scale=1]{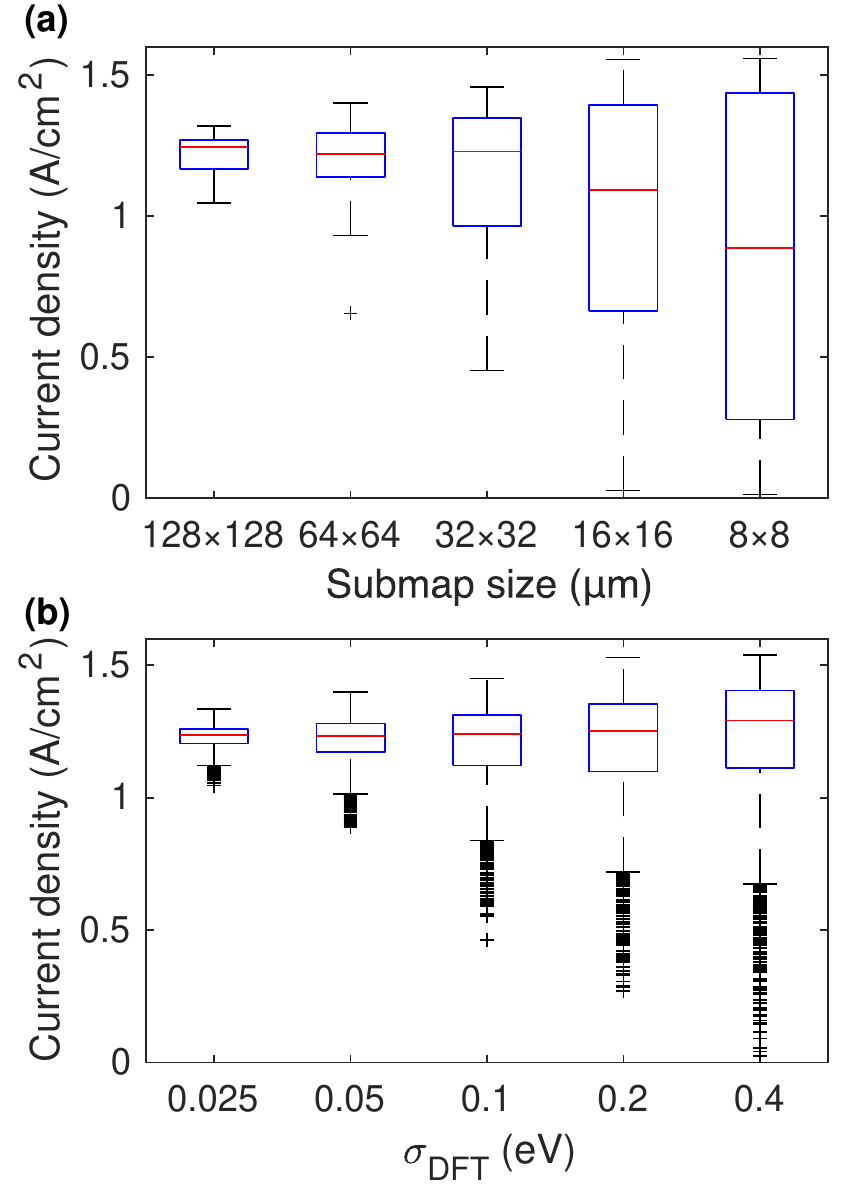}
	\caption{\label{p2f7}
		Boxplots showing the variability of the emitted current density predicted (a) from the submaps of different characterized area sizes at temperature $T=1250 \units{K}$, anode-cathode voltage $V = 400 \units{V}$ and distance $d=1.132 \units{mm}$, (b) from a work function map by applying $\phi(hkl)=\phi\subm{DFT}(hkl)+ \Delta\phi(hkl)$ to the grain orientation map in Fig. \ref{p2f3}a, where the shift values are independent and identically distributed (i.i.d.) following the normal distribution $\Delta\phi(hkl)\sim N(0.176\units{eV},\sigma\subm{DFT}^2)$. In a boxplot, the red line in the box indicates the median value. The lower edge of the box is the first quartile ($Q_1$ or 25th percentile). The upper edge is the third quartile ($Q_3$ or 75th percentile). The interquartile range (IQR) is defined as $\mathrm{IQR}=Q_3-Q_1$. Data points larger than $Q_3+1.5\mathrm{IQR}$ or smaller than $Q_1-1.5\mathrm{IQR}$ are considered as outliers and plotted individually using the $+$ symbols. The whiskers extend to the most extreme data points not considered to be outliers.
	}
\end{figure}

Fig. \ref{p2f7} shows the variability of the values of the emitted current density for different submap sizes (Fig. \ref{p2f7}a) and different uncertainties of work function values (Fig. \ref{p2f7}b) at a condition for the TL-FSCL transition region. Fig. \ref{p2f7}a shows how the prediction of the emitted current density becomes more precise as the size of the submap increases, which indicates that model users may determine the submap size to use according to their desired precision in the prediction. Fig. \ref{p2f7}b estimates the uncertainty in the predicted emitted current density as a function of the uncertainty in the work function values. Previous studies\cite{DeWaele2016,Tran2019} estimated that the error of the DFT work function values is on the scale of tenths of eV. Our results show that even in the extreme case that the DFT work function values have an uncertainty of $0.4\units{eV}$, the median (the red line in the box in Fig. \ref{p2f7}b) is $1.29 \units{A/cm^2}$, close to $1.23\units{A/cm^2}$, which was the result for the baseline case ($\phi=\phi\subm{DFT}+ 0.176\units{eV}$, Fig. \ref{p2f4} and \ref{p2f5}b). In the $\phi\subm{DFT}=0.4\units{eV}$ results, the first quartile (the lower edge of the box) is $Q_1=1.11\units{A/cm^2}$ while the third quartile (the upper edge) is $Q_3=1.41\units{A/cm^2}$, and the interquartile range is $\mathrm{IQR}=Q_3-Q_1=0.29\units{A/cm^2}$. Such a dispersion is smaller than using a $32\units{\mu m}\times32\units{\mu m}$ submap, which has $\mathrm{IQR}=0.37 \units{A/cm^2}$, indicating a robustly predicted average current density even for the higher end of DFT work function uncertainty values.

Other possible causes of the error in the predicted emitted current density include the measurement error in temperature values, the dependence of the work function value on temperature due to different stable arrangements of Ba-O surface species at different temperatures, and the dependence of the anode-cathode distance on temperature due to thermal expansion. The assumptions in the nonuniform emission model\cite{Chen2021} may also contribute to the error in the predicted emission, which include the assumption of a perfectly flat cathode surface in an infinite parallel diode, neglecting the lateral motion of the electrons, and neglecting the quantum effects. While it is beyond the scope of the present work to perform an in-depth investigation of the role of each of these sources of error, we find it very encouraging that the results in Fig. \ref{p2f4} illustrate that our model shows near quantitative agreement with experiment over a wide range of temperatures and anode-cathode voltages. This strong agreement with experiment suggests that while many sources of uncertainty in our model exist, they likely play a minor role in the resulting emission compared with the microstructural features of the cathode, including the fractions of each surface present, their size and spatial distribution, and the relative work functions of grains comprising the cathode surface.

\section{Conclusions and Outlook}
Our nonuniform emission model can predict two-dimensional maps of emitted current density and therefore the average emitted current densities at different temperatures and anode-cathode voltages based on a two-dimensional work function map derived by DFT calculations and microstructure characterization. Importantly, the predicted $J-T$ and $J-V$ curves are in semi-quantitative agreement with experimental results, including the nature of the TL-FSCL transition, which shows the same shape as experiments. There are only two fitting parameters in our model: the anode-voltage distance and a constant shift on the DFT work function values. The effect of these two fitting parameters on the shape of the TL-FSCL transition is negligible. Our model is the first of its kind to use a physics-based modeling method coupled with experimental characterization to reproduce experimental emission data, and illustrates that it is not necessary to use an empirical equation such as the Longo-Vaughan equation or to assume a continuous work function distribution. A key result of this work is that a smooth TL-FSCL transition region is a natural consequence of the physics of the nonuniform emission from a spatial distribution work function map when the effects of 3-D space charge, patch fields, and Schottky barrier lowering are included. 

The present findings provide both a robust physics-based approach to predict the emitted current from any polycrystalline cathode for which the surface grain orientations and work functions are known, and a means to understand how the cathode microstructure and the underlying work functions couple to the expected emission behavior. The present “forward” model starts from cathode work function distributions and predicts $J-T$ and $J-V$ curves. In the future, it may be possible to to create an “inverse” model where one starts from experimentally measured $J-T$ and/or $J-V$ curves and predicts an effective cathode work function arrangement and associated microsctructure consistent with the measured emission. Such an approach may be an effective method to better understand the coupling of cathode microstructure with the measured emission of new cathodes. Such a method would provide a powerful tool for understanding the expected emission behavior of new cathodes, as conducting an emission test on a new cathode is less time-consuming than a full suite of microstructure and work function studies, e.g., using EBSD characterization and DFT calculations. The results in this work can also be used as input for higher-level simulation codes like MICHELLE\cite{Petillo2002The-michelle-th} to improve the modeling of cathodes in electron gun fixtures, better informing device design and enabling deeper insight into the physical factors governing heterogeneous emission from thermionic cathodes. 

Codes for the nonuniform emission model are available on GitHub (https://github.com/chen-dongzheng/nonuniform-emission).

\section*{Acknowledgments}
This work was funded by the Defense Advanced Research Projects Agency (DARPA) through the Innovative Vacuum Electronic Science and Technology (INVEST) program with Leidos, Inc. The authors would like to thank Daniel Busbaher from 3M Technical Ceramics for providing the cathodes from which the data was obtained. The authors gratefully acknowledge use of facilities and instrumentation supported by NSF through the University of Wisconsin Materials Research Science and Engineering Center (DMR-1720415).

\bibliographystyle{IEEEtran}
\bibliography{Library}

\end{document}